\documentclass[aps,showpacs,superscriptaddress,twocolumn,amsmath,amssymb,prl]{revtex4}
\usepackage{dcolumn}% Align table columns on decimal point
\usepackage{bm}% bold math
\usepackage[dvips]{graphicx}
\usepackage{wrapfig,subfigure}
\usepackage{amssymb}
\usepackage{mathrsfs}
\usepackage{color}
\usepackage{ulem}
\usepackage{subfigure}
\begin{document}
\title{Spintronics-based mesoscopic heat engine}
\author{J. Atalaya and L. Y. Gorelik}
\affiliation{Department of Applied Physics, Chalmers University of Technology, G{\"o}teborg Sweden, SE-412 96}

\begin{abstract}
We consider a nanowire suspended between two spin-polarized leads and subject to a nonuniform magnetic field. We show that a temperature drop between the leads can significantly affect the nanowire dynamics. In particular, it is demonstrated that under certain conditions
the stationary distribution of the mechanical subsystem has Boltzmann form with an effective temperature which is smaller than the temperature of the "cold" lead; this seems rather counter-intuitive. We also find that a change of the direction of the temperature gradient results in generation of mechanical vibrations rather than heating of the mechanical subsystem.
\end{abstract}
\maketitle

Nanomechanical resonators are devices which are being employed not only to develop new technological applications, such as ultra-sensitive
sensors~\cite{LaHaye2004,Roukes2006,Rugar2004,Roukes1998}, but also to shed light on fundamental questions such as the transition from  the classical to the quantum mechanical description of macroscopic objects~\cite{Blencowe2004}. Investigation of a system where a mechanical degree of freedom controls the properties of a mesoscopic junction between two bulk leads is an important line of research in nanomechanics~\cite{HPark2000, LeRoy2004, Sazonova2004}. In such structures, the mechanical part may be considered as a nanoengine whose operation is controlled by the states of two bulk thermodynamic reservoirs. It is well known that a macroscopic mechanical system may be driven into cyclic motion if it is coupled to reservoirs  held at different temperatures. A Stirling engine, operating by cyclic compression and expansion of air or other gas and placed between hot and cold spaces, is one example.

By decreasing the size of the heat engine to the nanoscale level, quantum mesoscopic effects come into play and determine the behavior of both the working substance and the mechanical subsystem. This opens new possibilities for the operation of a heat engine; for instance, reduction or even suppression of the mechanical fluctuations---effective cooling of the mechanical subsystem. Recently, it was shown that suppression, leading to ground state cooling, may be achieved if reservoirs, presented by normal or superconducting metal leads, are held at different electrochemical potentials~\cite{Bachtold2009,Bachtold2010,Sonne2010,Fabio2011}. It was also demonstrated that a temperature drop between reservoirs can also generate this effect if one assumes a very special "three particles" interaction inside a junction~\cite{Mahler2010, Linden2010}.
In this letter we investigate heating,  pumping  and cooling of a mechanical mode in a realistic nanojunction where only a \emph{"two particle"} interaction between mechanical degree of freedom and working subsystem exists.  We show that a temperature drop between linked leads can generate cooling or excitation, depending on its direction, of the mechanical subsystem. 
\begin{figure}
\centering
\subfigure{
\includegraphics[width=8.0truecm,trim=0cm 4cm 0cm 4cm, clip=true]{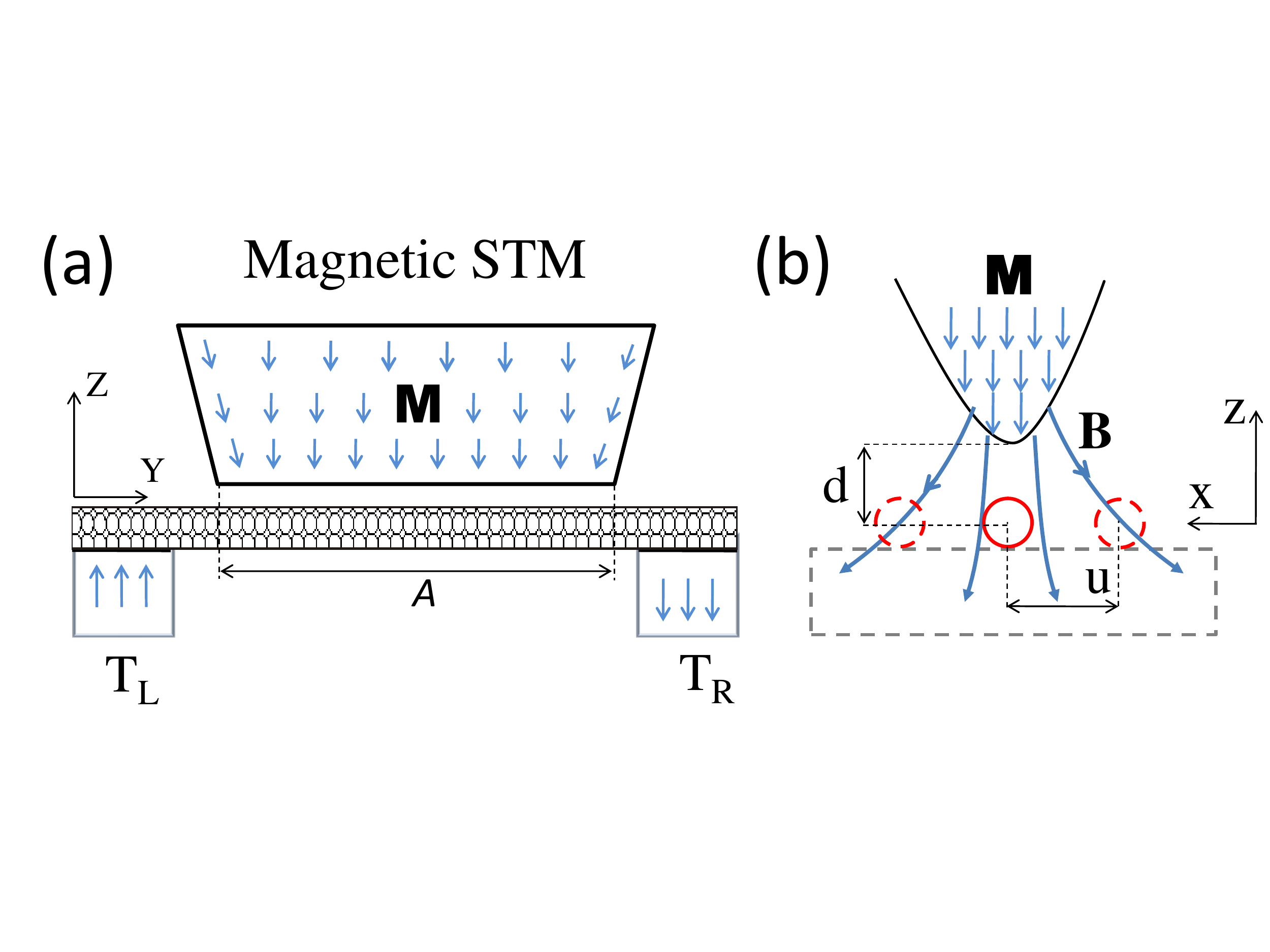}
\label{fig:fig1a}
}\\
\subfigure{
\includegraphics[width=8.0truecm,trim=0cm 1.5cm 0cm 7.8cm, clip=true]{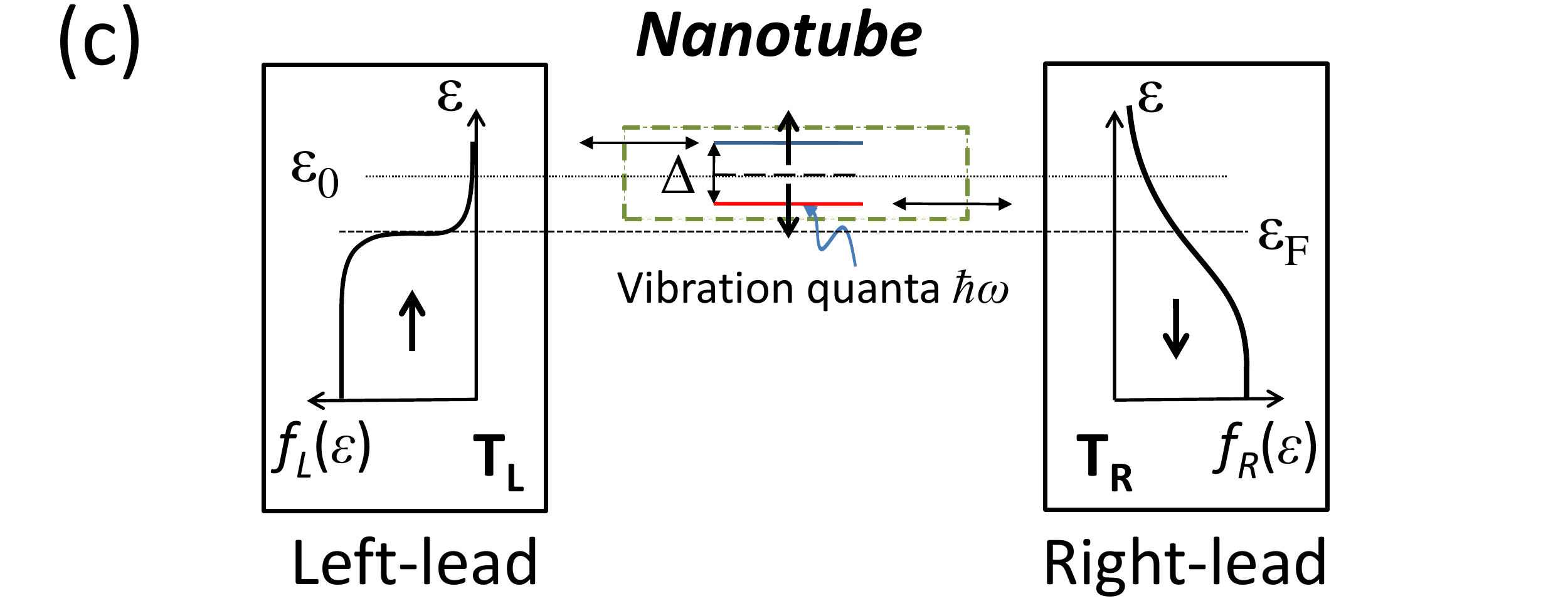}
\label{fig:fig1b}
}
\label{fig:fig1}
\caption{{\bf (a)} A nanotube suspended between two spin-polarized leads and in proximity to a magnetic STM tip with magnetization $\textbf{M}$. The leads have opposite polarizations along the $z$-direction. {\bf (b)}~Nonuniform magnetic field, {\bf B}, created by the magnetic tip. The nanotube~(circles) deflection,  $u$, is in the $x$-direction. {\bf (c)} A doubly spin-degenerate electronic level exists in the nanotube at energy $\epsilon_0$. The applied magnetic field splits this level into two levels $\sigma_z=\uparrow,\downarrow$ separated by an energy $\Delta\propto B_z\mu_{\textrm{B}}$. The leads are held at different temperatures $T_{L,R}$ with a zero bias voltage.  \vspace*{-3.0ex}  }
\end{figure}

To be specific, we consider a carbon nanotube suspended between La$_{1-x}$Sr$_{x}$MnO$_3$ ferromagnets of opposite polarizations (a structure recently realized  experimentally~\cite{Hueso2007}) and subject to a nonuniform magnetic field $\mathcal{\bf B}$ (cf.~Fig.~1b). The field can be generated, for example,  by a magnetic STM tip in the form of a wedge which is in proximity to the nanotube (cf.~Fig.~1a).  If the  nanotube is not too long (length $L\lesssim 1\mu$m), it can be considered as a quantum dot coupled to the electrodes through tunnel junctions. We assume that inside the nanotube there is only one doubly degenerate, with respect to spin, and spatially quantized electronic level which participates in electron exchange with the ferromagnetic leads.

If the nanotube is straight and positioned below the STM sharp end, as shown in Fig.~1b, the applied magnetic field is directed towards the nanotube ($z$-direction) and Zeeman-splits the degenerate electronic level of the nanotube. A two-level system (TLS) is thus formed in the latter with levels $\sigma_z=\uparrow,\downarrow$ and energies $\epsilon_{\uparrow,\downarrow}$, respectively. Deflection of the suspended part of the nanotube in the $x$-direction generates interlevel transitions (spin-flip) in the TLS. Consequently, the nonuniform magnetic field induces coupling between mechanical and electronic subsystems. In what follows, we will refer to this mechanism of interaction between mechanical and electronic subsystem as \textit{spintronic coupling}.

We assume that the left and right leads are completely polarized along the $z$-direction (\textit{i.e.}, the density of states are $\nu_{R(L)}^{\uparrow(\downarrow)}=0$ and $\nu_{R(L)}^{\downarrow(\uparrow)}\equiv\nu_{R(L)}$~$>0$) and the energy difference between levels of the TLS is $\Delta\equiv\epsilon_{\uparrow}-\epsilon_{\downarrow}>0$. 
If there is no spintronic coupling and the intrinsic relaxation time $\tau$ of the TLS is larger than the dwell-time of an electron in the nanotube, then the occupation number of the spin up (down) state inside nanotube is  $n_{\uparrow(\downarrow)}\simeq f_F\big((\epsilon_{\uparrow(\downarrow)}-\epsilon_{FL(R)})/T_{L(R)}\big)\equiv f_{L(R)}$, where $f_F$ is the Fermi-distribution function, $\epsilon_{FL(R)}$ and $T_{L(R)}$ are the Fermi energy and temperature of the left (right) lead, respectively.

The spintronic coupling generates spin-flip transitions between the energy levels of the TLS. These transitions are inevitably accompanied by absorption or emission of mechanical quanta. The cooling~(pumping) process of the mechanical subsystem is the result of an electron transition from the lower~(upper) energy level to the upper~(lower) energy level of the TLS. In order to effectively cool~(pump) the mechanical mode, it is required that $n_{\uparrow}\ll n_{\downarrow}$ ($n_{\uparrow}\gg n_{\downarrow}$). This can be realized by applying a bias voltage or by subjecting the leads to different temperatures. Below, we consider cooling and pumping of the mechanical subsystem generated only by a temperature gradient between the leads.

To perform a quantitative analysis of the system described above, we consider the  Hamiltonian
\begin{equation}
H = H_{nw} + H_{l} + H_{t} , \label{eq:Hamiltonian}
\end{equation}
\begin{eqnarray}
H_{nw} &=& \hbar\omega \hat{b}^{\dag}\hat{b}  + \sum_{\sigma=\uparrow,\downarrow}\epsilon_{\sigma} \hat{n}_{\sigma} +  g\hat{u}(\hat{d}_{\uparrow}^{\dagger}\hat{d}_{\downarrow}+\hat{d}_{\downarrow}^{\dagger}\hat{d}_{\uparrow})\label{eq:H_NW}, \\
H_{l} &=& \sum_{k} \epsilon_{L}(k) \hat{a}^\dagger_{k,\uparrow, L }\hat{a}_{k,\uparrow, L} + \epsilon_R(k) \hat{a}^\dagger_{k,\downarrow, R}\hat{a}_{k,\downarrow, R}, \label{eq:H_L}\;\;\;\; \\
H_{t} &=& \sum_{k}t_{L}\hat{a}^\dagger_{k,\uparrow,L}\hat{d}_{\uparrow} + t_{R}\hat{a}^\dagger_{k,\downarrow, R}\hat{d}_{\downarrow} + h.c.,  \label{eq:H_T}
\end{eqnarray}
where $\hat{a}_{k,\sigma,L(R)}(\hat{a}_{k,\sigma,L(R)}^{\dag})$ and $\hat{d}_{\sigma}(\hat{d}_{\sigma}^{\dag})$
being the annihilation (creation) operators for
electrons in the left (right) leads and in the nanotube, respectively, and $ \hat{n}_{\sigma} =\hat{d}_{\sigma}^{\dagger}\hat{d}_{\sigma}$.
The first term in Eq.~\eqref{eq:H_NW}  describes the nanotube mechanical degrees of freedom, which we restrict
to the fundamental flexural mode. This mode is described as a
simple harmonic oscillator with vibrational frequency $\omega$ and $\hat{b}(\hat{b}^{\dag})$
being the annihilation (creation) operator for an elementary
excitation (vibron). The second term in Eq.~\eqref{eq:H_NW} describes the TLS with energy levels
$\epsilon_{\sigma}=\epsilon_0\pm\Delta/2$, where $\Delta$ is the Zeeman-splitting energy proportional to the $z$-component of the applied magnetic field $B_z$, and $\epsilon_0$ is the zero-field energy, which is measured relative to the leads Fermi energy.
The last term in Eq.~\eqref{eq:H_NW} describes the spintronic
coupling. It is proportional to the oscillator displacement, $\hat{u}=(\hat{b}^{\dag}+\hat{b})/\sqrt{2} $,  and to the spin-flip operator. The coupling parameter  $g$ is equal to $C\mu_{\textrm{B}} x_{0} \partial_x B_x(\textbf{0})$ where $\mu_{\textrm{B}}$ is the Bohr magneton, $x_{0}$ is the zero-point vibrational amplitude, $\partial_xB_x(\textrm{\bf 0})$ is the field gradient along the $x$-direction, and $C$ is a numerical factor $ \sim A/L$, where $L$ is the nanowire length and $A $ is the length of the STM wedge. This factor accounts for the electronic state inside the nanotube being extended over the whole length of the nanotube while the magnetic field is concentrated only in the region below the STM.  The field gradient $\partial_x B_x(\textbf{0})$ induced by a Fe-based magnetic tip 7.5~nm thick with magnetization $M=1.75~\cdot 10^{6}$~A/m at a distance 7.5~nm is of the order of $35$~mT$\cdot$nm$^{-1}$. For this value of field gradient and for a nanotube with vibrational frequency $\omega=2\pi\cdot100$~MHz, mass 1~ag and $A/L=0.1$, $g$ is of the order of $2\pi\cdot 10^6$~Hz. The term $H_l$ in Eq.~\eqref{eq:H_L} describes the left and right leads.  The term $H_t$ in Eq.~\eqref{eq:H_T} describes the tunneling of electrons from the nanowire to the leads, and \textit{v.v.}, and $t_{L(R)}$ are tunneling amplitudes~\cite{remark1}.

To analyze the performance of the system, we  start from the Liouville-von Neumann equation for the total density
operator $\hat{\varrho}$ and  then eliminate the lead electronic degrees of freedom~\cite{Gorelik2005}. If  the temperatures of the leads are much greater than $\hbar\omega/k_{\textrm{B}}$, the Fermi-distributions $f_{L,\,R}(\epsilon)$ are smooth  functions within the energy interval $\hbar\omega$.  As a result one gets  the following Lindblad master equation~\eqref{eq:MEq} for the reduced density matrix $\rho= \mathrm{Tr}_{R+L}\hat{\varrho}$. The latter describes the mechanical degree of freedom and the electronic state of the TLS of the nanotube.
\begin{equation}
\partial_t\rho = -\frac{i}{\hbar}\big[H_{nw},\rho\big] + \mathcal{L}_{L}[\rho] + \mathcal{L}_{R}[\rho], \label{eq:MEq}
\end{equation}
 where  
\begin{eqnarray}
\mathcal{L}_{\alpha}[\rho] &=& \Gamma_{\alpha}[(1-f_{\alpha})\hat{d}_{\sigma_{\alpha}} \rho \hat{d}_{\sigma_{\alpha}}^{\dagger}
 + f_{\alpha}(\hat{d}_{\sigma_{\alpha}}^{\dagger} \rho \hat{d}_{\sigma_{\alpha}} \rho)- \nonumber\\
&& - (1/2-f_{\alpha})\{ \hat{n}_{\sigma_k},\rho\}].
\end{eqnarray}
Here, $\alpha=(L,R)$, $\sigma_{L(R)}=\uparrow(\downarrow)$ and $\Gamma_{\alpha}=2\pi|t_{\alpha}|^{2}\nu_{\alpha}/\hbar$ are the tunneling rates, $\nu_{\alpha}$ is the density of states, and $\{\hat{A},\hat{B}\}$ denotes anticommutator. The collision integrals $\mathcal{L}_{\alpha}[\rho]$ describe the decoherence in the electronic subsystem induced by the bulk electronic reservoirs.

If the resonant condition $\Delta=\hbar\omega$ is fulfilled  and $\omega \gg g/\hbar$, $\Gamma_{L}$, the rotating-wave-approximation  can be used to obtain the following rate equations
\begin{eqnarray}
\dot{P}_m(n) &=& \tilde{g}\big( P^i_{\uparrow\downarrow}(n) - P^i_{\uparrow\downarrow}(n+1)\big), \label{eq:RWA} \\
\dot{P}_{\uparrow}(n) &=& -\tilde{g}P^i_{\uparrow\downarrow}(n+1) - \Gamma_R f_R P_{\uparrow}(n)+ \Gamma_Lf_LP_0(n) \nonumber \\
&&  - \Gamma_L(1-f_L)P_{\uparrow}(n) + \Gamma_R(1-f_R)P_2(n) ,\nonumber\\
\dot{P}_{\downarrow}(n)  &=&  \tilde{g}P^i_{\uparrow\downarrow}(n)+ \Gamma_R f_R P_0(n) -\Gamma_L f_L P_{\downarrow}(n)  \nonumber \\
&& + \Gamma_L(1-f_L)P_2(n)  - \Gamma_R(1-f_R)P_{\downarrow}(n), \nonumber\\
\dot{P}_2(n) &=& - \Gamma_L(1-f_L)P_2(n) -\Gamma_R(1-f_R)P_2(n)   \nonumber \\
&&+\Gamma_L f_L P_{\downarrow}(n) +\Gamma_R f_R P_{\uparrow}(n),\nonumber \\
2\dot{P}^i_{\uparrow\downarrow}(n) &=& \tilde{g}n\big(P_{\uparrow}(n-1) -  P_{\downarrow}(n) \big) - (\Gamma_L+\Gamma_R) P^i_{\uparrow\downarrow}(n),\nonumber
\end{eqnarray}
where $\tilde{g}=g\sqrt{2}/\hbar$. Here $P_{\sigma}(n)$, $P_0(n)$, $P_{2}(n)$  are the joint probabilities to find a vibrational mode in a Fock state with $n$ vibronic quanta and one electron on the nanotube with spin $\sigma$, empty nanotube and two electrons on the nanotube, respectively. Therefore, the total probability to find $n$ vibronic quanta is  $P_{m}(n)\equiv P_{0}(n)+P_{\uparrow}(n)+P_{\downarrow}(n)+P_{2}(n)$. The off-diagonal elements of the density matrix $P^i_{\uparrow\downarrow}(n)=\textrm{Im}\langle 0|\hat{d}_{\uparrow}b^{(n-1)}\rho (\hat{b}^{\dag})^{n}\hat{d}^{\dag}_{\downarrow}|0\rangle/(n-1)!$ describe the quantum entanglement between the electronic and mechanical subsystems generated by the correlation between  spin-flip of electron and a change in the number of  vibronic quanta.
\begin{figure}[t]
\includegraphics[width=\linewidth,trim=0cm 3cm 0cm 2cm, clip=true]{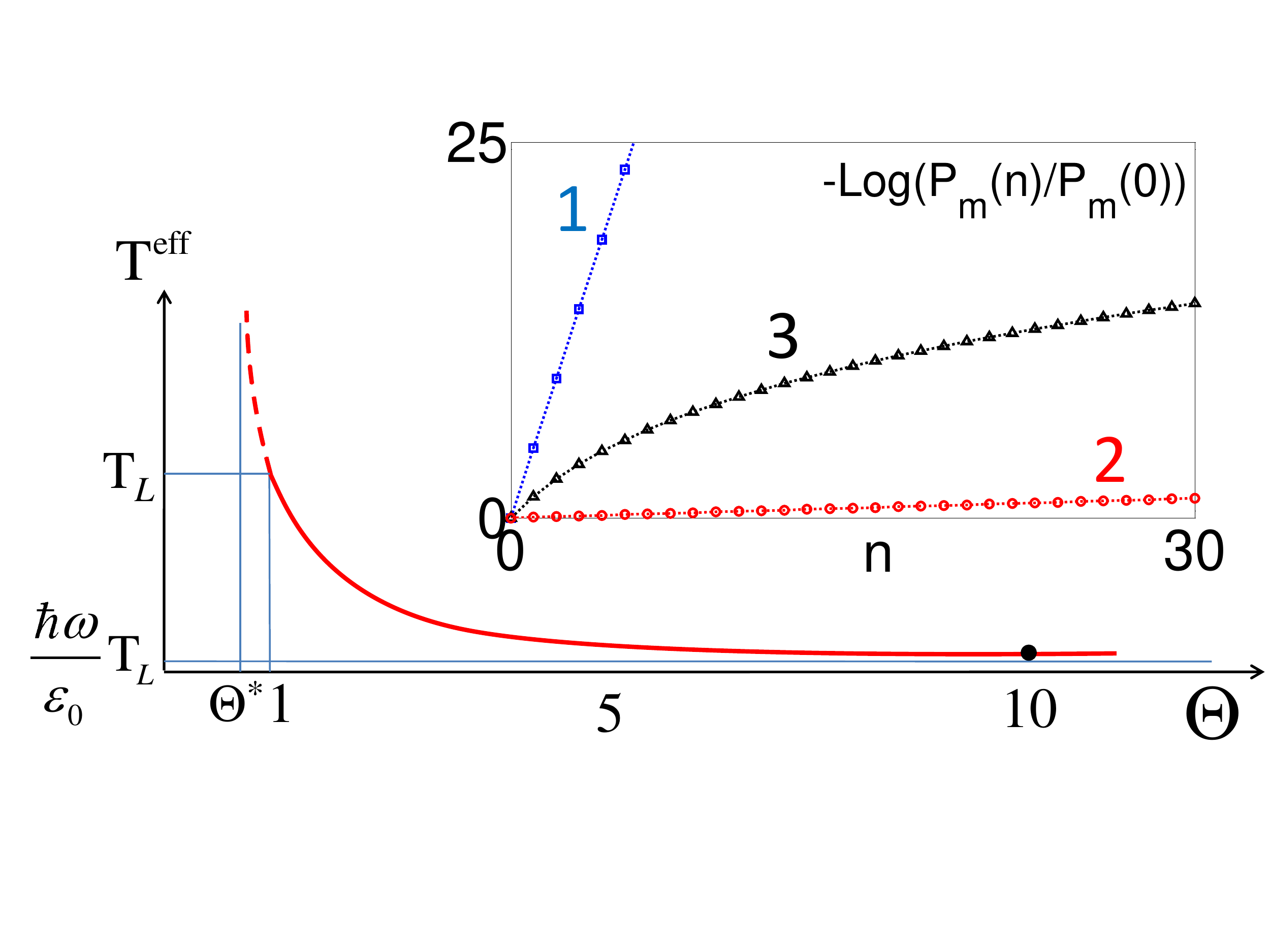}
\caption{Effective temperature $T^{eff}$ of the vibrational mode  as function of the ratio $\Theta=T_R/T_L$. For $\Theta>1$: $T^{eff}<\min\{T_L,T_R\}$. For $\Theta^*<\Theta<1$:  $T^{eff}> \max\{T_L,T_R\}$. For $\Theta<\Theta^*$: no stationary distribution exits unless additional dissipation mechanisms are included. {\bf Inset:} Vibron stationary distribution in logarithmic scale when the mechanical subsystem interacts only with ferromagnetic leads (curve 1), only with a bosonic bath at temperature $T_b$ and coupling parameter $\gamma$ (curve 2) and, both the fermionic and bosonic  baths (curve 3).  We use $\omega/2\pi=100$~MHz, $T_L=0.02$~K, $\Theta=10$, $T_b=(T_L+T_R)/2$,  $\epsilon_0= k_{\textrm{B}}T_R/2$, $\Gamma_L=\sqrt{2}g/\hbar$, $\Gamma_R=1.62g/\hbar$ and $\gamma=0.002g/\hbar$. \label{fig:fig2}}\vspace*{-3.0ex}
\end{figure}

Equation~\eqref{eq:RWA} always has a stationary solution where $P^{st}_m(n)$ has a Boltzmann form
\begin{equation}
P^{st}_m(n) = Z^{-1}\exp\big(-\hbar\omega n/k_BT^{eff}\big), \label{eq:Pst}
\end{equation}
where $Z=\big[1-\exp(-\hbar\omega/k_{\textrm{B}}T^{eff}) \big]^{-1}$ and the effective temperature is given by
\begin{equation}
 T^{eff}= T_L\frac{\hbar\omega}{\epsilon_0} \Big[1-\Theta^{-1} + \frac{\hbar\omega}{2\epsilon_0}\big( 1+ \Theta^{-1}\big)\Big]^{-1}, \label{eq:Teff}
\end{equation}
where $\Theta=T_R/T_L$. This solution has physical meaning only if $T^{eff}>0$. Negative effective temperature indicates that there is permanent energy pump into the mechanical subsystem. In this case in order to achieve a stationary regime one has to introduce additional external sources of dissipation. 

From Eq.~\eqref{eq:Teff} one can see that for $\epsilon_{\downarrow}>0$ and $T_R>T_L$, the vibrational mode is effectively cooled to a final temperature which is smaller than the temperature of the "cold" lead ($T^{eff}<\min\{T_L,T_R\}=T_L$)---cooling regime. From this equation also follows that in the  interval $1>\Theta>\Theta^*\equiv\big(1-\hbar\omega/2\epsilon_0\big)/\big(1+\hbar\omega/2\epsilon_0 \big)$
the effective temperature is greater than the temperature of the "hot" lead ($T^{eff}>\max\{T_L,T_R\}=T_L$)---heating regime.

Note that when a temperature difference is used to reduce mechanical fluctuations of the nanotube, the minimum effective temperature of the mechanical subsystem is always greater than $T^{eff}_{min}=2T_L/(1 +2\epsilon_0/\hbar\omega)$, approaching this value as $\Theta\rightarrow \infty$. As a result, the average vibron number $\langle n\rangle=\sum_{n}nP_m(n) \sim\exp(-\epsilon_0/k_BT_L)$ is finite but exponentially small. This is a consequence of the finite temperature of the "cold" lead which leads to infrequent  pumping processes and does not allow to achieve absolute ground state cooling.
\begin{figure}
\includegraphics[width=0.9\linewidth,trim=0.4cm 1.0cm 0.2cm 2.8cm, clip=true]{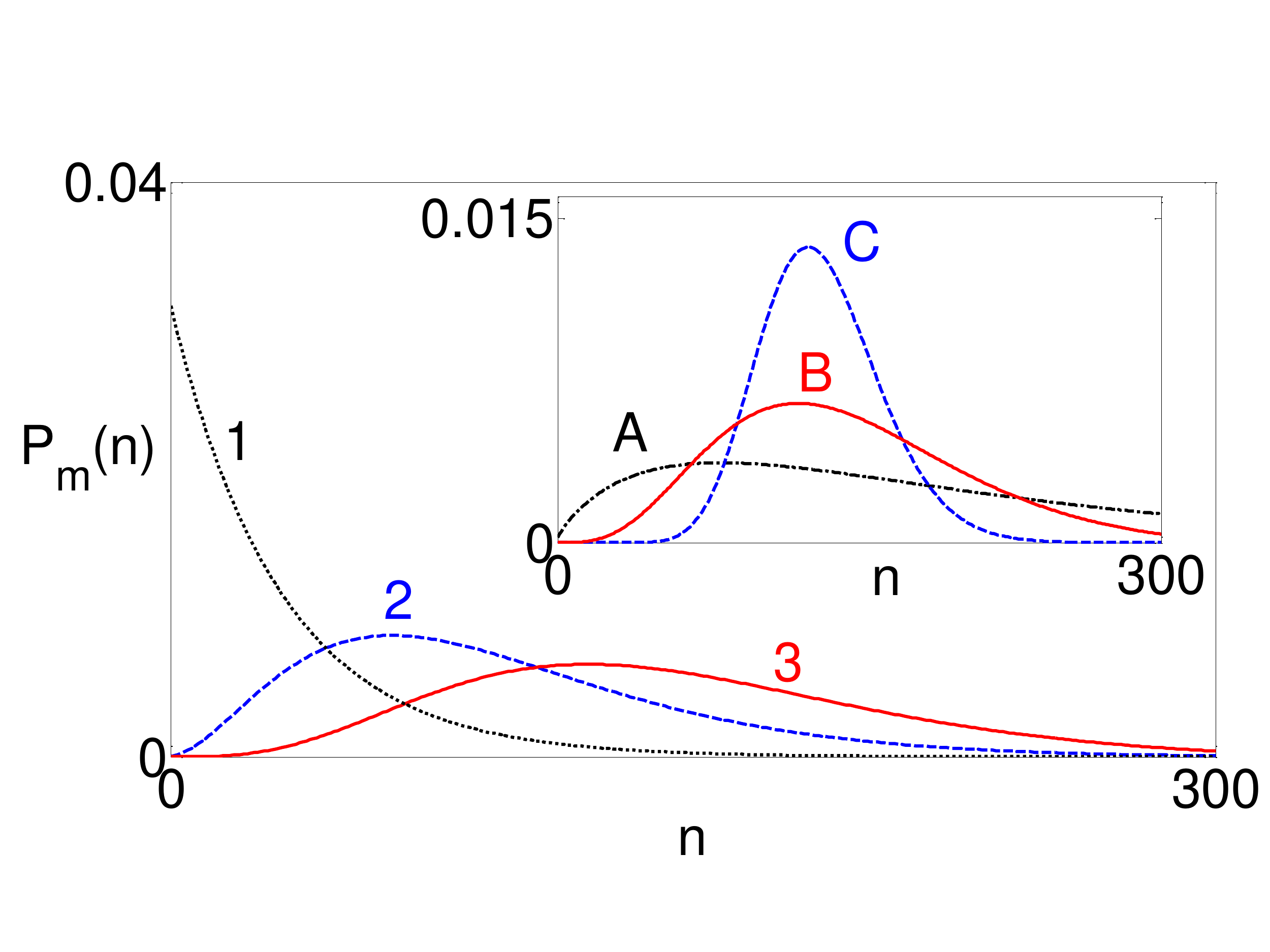}	
\caption{Vibron stationary distribution $P_m(n)$ when $\Theta<\Theta^*$. We include a bosonic bath with temperature $T_b$ and coupling parameter $\gamma$. The stationary distribution exhibits a peak at $n_{max}$, which increases as $\gamma$ gets smaller. We use $n_b\gamma\hbar/g\gg1$ (curve 1), $n_b\gamma\hbar/g = 0.021$ (curve 2) and $n_b\gamma\hbar/g = 0.01$ (curve 3). {\bf Inset:} The width of the distribution $P_m(n)$ scales inversely with $T_b$.  $T_b(A)=1.5$~K, $T_b(B)=0.3$~K, $T_b(C)=0.06$~K and $n_b\gamma\hbar/g = 0.01$ for all curves A, B and C. \label{fig:fig3}}
\end{figure}

From Eq.~\eqref{eq:RWA} with $T_L$ and $T_R$ such that $T^{eff}<0$ ($\Theta<\Theta^*$), it follows that $\textrm{d}\langle n \rangle/\textrm{d}t>0$ and the average number of vibrons increases with time. However, in a real physical situation the mechanical subsystem is also coupled to the phononic thermal baths of the leads. To account for the dissipation due to this coupling we add to the left side of Eq.~\eqref{eq:MEq} a Lindblad operator $\gamma\mathcal{L}_{\gamma}$~\cite{HPBreuer2002}. Here, the temperature of the phononic bath is $T_b\sim\big(T_L+T_R\big)/2$ and $\gamma=\omega/Q$ is the coupling parameter characterized by the resonator quality factor $Q$, which can be very large $Q\simeq 10^{5}$  at low temperatures~\cite{Huttel2009}.  We  properly modify the system of equations~\eqref{eq:RWA},  and solve it numerically. The resulting shape of $P^{st.}_m(n)$ is depicted in Fig.~\ref{fig:fig3}. From this figure one can see that at small $n$ the probabilities $P^{st.}_m(n)$ increase with $n$. This is because for small $n<\Gamma/g$  and $\gamma < g$, the pumping rate ($\propto gn$) is larger than the dissipation rate ($\propto\gamma n$). Then, the pumping rate eventually saturates to $\Gamma\simeq\min\Gamma_{L(R)}$ at $n = n_s \sim \Gamma/g$ while the  the dissipation rate governed by interaction with phonon reservoirs continue to increase linearly with $n$ and finally overcomes the pumping rate. As a result, the distribution function reaches a maximum at $n=n_{max}\sim\Gamma/\gamma$ and then exponentially decays for larger $n\gg n_{max}$.

Finally, we discuss conditions for ground state cooling of the mechanical subsystem; \textit{i.e.,} a stationary regime with a final vibron number $\langle n \rangle \lesssim1$. Thus far, we have considered the situation of completely polarized leads. However, in order to analyze the conditions for ground state cooling, it is necessary to estimate the effect of partial spin-polarization in the  leads. To quantify the degree of polarizability of the right lead, we introduce a parameter $\eta_{L(R)} = 1-\nu_{L(R)}^{\downarrow(\uparrow)}/\nu_{L(R)}^{\uparrow(\downarrow)}$ and find that, for a symmetric case ($\Gamma_{R}=\Gamma_{L}$), $1-\eta_R \ll1$, $T_L\ll T_R$ and $\epsilon_0\sim T_Rk_{\textrm{B}}\gg\hbar\omega$, the average number of vibrons is 
\begin{equation}
\langle n \rangle \approx \langle n \rangle_{\eta=0} +(1-f_{R})(1-\eta_R) + \mathcal{O}\big((1-\eta_R)^2\big).
\end{equation}
Hence, for a 90\% spin-polarized right lead, the average vibron number is increased only by $\approx 0.1$.

A necessary condition to achieve ground state cooling comes from Eq.~\eqref{eq:Teff} which suggests that in order to achieve the minimum effective temperature one needs $T_R\gg T_L$ and $\epsilon_0\gg\hbar\omega$. The sufficient condition comes from requirement that the cooling rate $\kappa$ should be larger than the damping rate $\gamma$ due to a bosonic bath. The largest cooling rate for a given temperature gradient is achieved when $\Gamma_L\gtrsim g/\hbar \gtrsim \Gamma_R(1-f_R)$. Thus, the conditions for ground state cooling of the mechanical mode in the presence of a bosonic bath are $T_R\gg T_L$, $\epsilon_0\gg\hbar\omega$ and $Q\gg \hbar\omega n_b/gf_{R}$, where $n_b=[\exp(\hbar\omega/k_{\textrm{B}}T_b)-1]^{-1}$ ($T_b$ is the temperature of the bosonic bath). From the last inequality, it follows that the best cooling regime is achieved when  $k_{\textrm{B}}T_{R}\approx\epsilon_{0}$. Our analysis shows that for a nanotube with frequency $\omega=2\pi\cdot 100$~MHz ($\hbar\omega/k_{\textrm{B}}\approx 6$~mK) and realistic  coupling parameter $g\approx 2\pi\cdot 10^6$~Hz (see above) and quality factor $Q=10^5$, the average vibronic number can be reduced to  $\langle n \rangle=0.44$ for $T_{R}$=200~mK and $T_{L}$=20~mK.

In conclusion we have studied heating, pumping and cooling of the mechanical vibrations of a nanotube suspended between two highly polarized magnetic leads. We have shown that spintronic coupling between the mechanical and electronic subsystems generated by a nonuniform magnetic field may result in suppression or generation of mechanical vibrations  when the leads are held at different temperatures. In particular, it was demonstrated that under certain conditions the stationary distribution of the mechanical subsystem has Boltzmann form with effective temperature which is smaller than the temperature of the "cold" lead. This counter-intuitive result is a consequence of the Fermionic nature of the thermal baths coupled to mechanical subsystem. Notice that coupling to bosonic baths results in an effective  temperature of the vibrational mode equal to the mean value of the baths temperature. Also changing direction of the temperature gradient results in generation of mechanical vibrations rather than heating of the mechanical subsystem.  Finally, for partial spin polarization in the leads, ground state cooling of the mechanical vibration can be achieved at realistic physical parameters if the leads have $\gtrsim 50$\% spin polarization.

This work was supported in part by the Swedish VR and SSF
and by the EC project QNEMS (FP7-ICT-233952). We also acknowledge Dr. Andreas Isacsson for helpful discussions.

%\bibliographystyle{apsrev}
%\bibliographystyle{apsrev4-1}
%\bibliography{ja2011,md10,JA2010}

\end{document}